\PassOptionsToPackage{unicode}{hyperref}
\PassOptionsToPackage{hyphens}{url}
\documentclass[
]{article}
\usepackage{lmodern}
\usepackage{amssymb,amsmath}
\usepackage{ifxetex,ifluatex}
\ifnum 0\ifxetex 1\fi\ifluatex 1\fi=0 
  \usepackage[T1]{fontenc}
  \usepackage[utf8]{inputenc}
  \usepackage{textcomp} 
\else 
  \usepackage{unicode-math}
  \defaultfontfeatures{Scale=MatchLowercase}
  \defaultfontfeatures[\rmfamily]{Ligatures=TeX,Scale=1}
\fi
\IfFileExists{upquote.sty}{\usepackage{upquote}}{}
\IfFileExists{microtype.sty}{
  \usepackage[]{microtype}
  \UseMicrotypeSet[protrusion]{basicmath} 
}{}
\makeatletter
\@ifundefined{KOMAClassName}{
  \IfFileExists{parskip.sty}{%
    \usepackage{parskip}
  }{
    \setlength{\parindent}{0pt}
    \setlength{\parskip}{6pt plus 2pt minus 1pt}}
}{
  \KOMAoptions{parskip=half}}
\makeatother
\usepackage{xcolor}
\IfFileExists{xurl.sty}{\usepackage{xurl}}{} 
\IfFileExists{bookmark.sty}{\usepackage{bookmark}}{\usepackage{hyperref}}
\hypersetup{
  hidelinks,
  pdfcreator={LaTeX via pandoc}}
\date{}

\usepackage{graphicx}
\usepackage{hyperref}
\usepackage{pdflscape} 
\usepackage{multirow}
\usepackage{setspace}
\usepackage{lineno}
\usepackage{authblk}
\usepackage{soul}

\usepackage{pdfpages}

\usepackage{fancyhdr}

\usepackage{booktabs}

\title{Using graph theory and social media data to assess cultural ecosystem
	services in coastal areas: Method development and application\footnote{© 2020. This manuscript version is made available under the CC-BY-NC-ND 4.0 license \href{http://creativecommons.org/licenses/by-nc-nd/4.0/}{http://creativecommons.org/licenses/by-nc-nd/4.0/}}}

\author[1]{Ana Ruiz-Frau}

\author[1]{Andres Ospina-Alvarez\footnote{Corresponding author: Andrés Ospina-Alvarez, email: aospina.co@me.com; address: Spanish Scientific Research Council, Mediterranean Institute for Advanced Studies (IMEDEA-CSIC/UIB), C/ Miquel Marques 21, CP 07190 Esporles, Balearic Islands, Spain.}}

\author[2, 3]{Sebastián Villasante}

\author[2, 3]{Pablo Pita}

\author[4]{Isidro Maya-Jariego}

\author[5]{Silvia de Juan Mohan}

\affil[1]{Mediterranean Institute for Advanced Studies (IMEDEA-CSIC/UIB), C/
	Miquel Marques 21, CP 07190 Esporles, Balearic Islands, Spain.}
\affil[2]{Faculty of Economics and Business Administration,
	University of Santiago de Compostela, Av~Burgo das Nacions~s/n, 15782
	Santiago de Compostela, A Coruña, Spain.}
\affil[3]{Campus Do Mar, International Campus of Excellence,
	Spain.}
\affil[4]{Department of Social Psychology, Universidad de
Sevilla, Calle Camilo José Cela s/n, 41018 Seville, Spain.}
\affil[5]{Marine Science Institute (ICM-CSIC), Passeig Marítim de la
	Barceloneta, 37-49, CP 08003 Barcelona, Catalunya, Spain.}

\begin{document}
	\maketitle

\textbf{Running page head:} Graph theory to assess CES from social media 

\begin{abstract}

The use of social media (SM) data has emerged as a promising tool for
the assessment of cultural ecosystem services (CES). Most studies have
focused on the use of single SM platforms and on the analysis of photo
content to assess the demand for CES. Here, we introduce a novel
methodology for the assessment of CES using SM data through the
application of graph theory network analyses (GTNA) on hashtags
associated to SM posts and compare it to photo content analysis. We
applied the proposed methodology on two SM platforms, Instagram and
Twitter, on three worldwide known case study areas, namely Great Barrier
Reef, Galapagos Islands and Easter Island. Our results indicate that the
analysis of hashtags through graph theory offers similar capabilities to
photo content analysis in the assessment of CES provision and the
identification of CES providers. More importantly, GTNA provides greater
capabilities at identifying relational values and eudaimonic aspects
associated to nature, elusive aspects for photo content analysis. In
addition, GTNA contributes to the reduction of the interpreter's bias
associated to photo content analyses, since GTNA is based on the tags
provided by the users themselves. The study also highlights the
importance of considering data from different social media platforms, as
the type of users and the information offered by these platforms can
show different CES attributes. The ease of application and short
computing processing times involved in the application of GTNA makes it
a cost-effective method with the potential of being applied to large
geographical scales.{~}
\end{abstract}


\section{Introduction}

Humans are deeply connected to the largest biome of the planet---the
Ocean. For centuries, humans have lived in coastal communities where
people fished, gleaned and hunted to support their livelihoods
(Erlandson and Rick, 2010). Humans are highly dependent on the benefits
and services provided by ecosystems (McMichael et al., 2005). Living by
the coast shapes cultures and identities whose actions influence the
marine and coastal physical environments to which coastal communities
are connected to (Klain et al., 2014). Marine and coastal ecosystem
services (ES), such as food provision, climate regulation or the
creation of opportunities for recreation and relaxation, are fundamental
elements in the maintenance of human wellbeing (Selig et al., 2019).
Human interactions with coasts can also affect mental health in many
ways, and the forms of evidence include positive effects related to
happiness, social interactions, social cohesion and engagement; a sense
of meaning and purpose in life; and decreases in mental distress
(Bratman et al., 2019).

Cultural ecosystem services (CES) are some of the benefits people can
most directly relate to, since most human-nature interactions fall
within the CES category (Garcia Rodrigues et al., 2017; Leenhardt et
al., 2015). However, marine ecosystem services {[}and CES in
particular{]} have been impacted at unprecedented rates by climate
change (namely in the form of ocean warming, ocean acidification,
deoxygenation, and sea level rise) and direct anthropogenic activities
(e.g. fishing, pollution and habitat degradation) (IPBES, 2019). In
addition, CES are often overlooked in conservation and management
schemes for marine and coastal areas (Chan et al., 2012; Everard et al.,
2010; Garcia Rodrigues et al., 2017). Defining human-nature interactions
in coastal areas and the type of CES offered, e.g. the activities people
undertake, what they value, what habitats or species attract most
attention, at scales relevant for marine and coastal management is time
consuming and often requires resources which are not generally available
(Waldron et al., 2013). In recent years, social media data, that is,
data created and shared by users on social media platforms, has emerged
as a potential useful source of information in environmental research,
management and conservation (Di Minin et al., 2015; Ghermandi and
Sinclair, 2019; Toivonen et al., 2019). Among the most popular social
media networking sites we find Facebook, YouTube, Instagram, Twitter or
Flickr (Di Minin et al., 2015; Toivonen et al., 2019). Typically, social
media users share data in the form of tags, text, images or videos
depending on the platform of choice. As an example, Instagram users
generally share images often complemented with a short text and relevant
tags selected by the user. In comparison, Twitter works as a
micro-blogging platform, where users share short messages (currently
limited to 270 characters) sometimes accompanied by an image. In
addition to differences in data content type, there are also differences
in users' demographic characteristics between social media platforms,
e.g., the proportion of females, young adults and teenagers is higher in
Instagram than in Twitter (PRC, 2019). Despite differences in data
content and user types, social media data mining and analysis has proven
very valuable as it can provide information on how people interact with
their environment, including interactions with nature (Di Minin et al.,
2015; Mancini et al., 2018) people's preferences for nature-based
experiences (Hausmann et al., 2017; Oteros-Rozas et al., 2018),
visitation patterns in conservation areas (Tenkanen et al., 2017; Wood
et al., 2013) or on mapping CES (Clemente et al., 2019; Richards and
Friess, 2015). So far, however, while the number of studies focusing on
the terrestrial environment is increasing, few had marine and coastal
areas within their scope (Ghermandi and Sinclair, 2019; Toivonen et al.,
2019).

Generally, social media data mining studies have restricted their scope
to a single social media platform (Ghermandi and Sinclair, 2019;
Toivonen et al., 2019), therefore limiting their assessments to
particular data formats, to certain sectors of the population (PRC,
2019), or to particular user's needs and behaviours (Manikonda et al.,
2016; Tenkanen et al., 2017). Logically, most studies have relied on
social media platforms that offer easy data access such as Flickr.
Flickr, a social media platform popular among nature photographers with
over 90 million monthly active users (2018), allows unrestricted access
to user posted content for non-commercial use through their Application
Programming Interface (API). On the other hand, Instagram, the most
popular social media platform (1 billion monthly active users in 2018)
after Facebook (2.26 billion users) (Ortiz-Ospina, 2019), has
increasingly restricted content access through their API since 2016. As
a consequence, the majority of studies have relied on Flickr as a source
of data (Ghermandi and Sinclair, 2019; Toivonen et al., 2019). However,
due to limited user numbers and post frequency, the amount of
observations provided by Flickr is sometimes too low to adequately
represent visitor rates in natural areas, as opposed to the higher
representativeness achieved through the use of Instagram (Tenkanen et
al., 2017). In addition, Flickr predominantly contains nature and
wildlife photography, while pictures including people are more frequent
in Instagram (Tenkanen et al., 2017). Therefore, limiting analysis to
Flickr data could lead to an over-representation of particular CES (e.g.
wildlife observation) while under-representing others, such as people
actively engaging with nature through an activity (e.g. recreational
activities).{~}

Regarding the methodological approaches used in the analysis of social
media data, a high proportion of studies have used images as a primary
source of information to assess the benefits associated to an area and
their spatial distribution (Wood et al., 2013). Geolocation of post
images has been used to assess the spatial distribution of the supply
and demand of CES (e.g. Clemente et al., 2019), while photo content
analysis provides information of the type of CES provided by a
particular area. Most studies have relied on the manual classification
of photo content; however, this is extremely time consuming. Recently,
new methodologies based on artificial intelligence and deep-learning
approaches have increasingly facilitated the automatic description of
photo content (e.g. Lee et al., 2019), reducing data processing time to
a fraction. While the application of artificial intelligence represents
a milestone in the analysis of photo content, it still presents some
challenges in its application and outcomes (Lee et al., 2019).{~}

To advance in the assessment of CES provided by nature through social
media data, we present a novel methodology based on the analysis of text
information associated to social media posts (i.e. hashtags) through the
application of graph theory network analysis techniques. Graph theory is
defined as the mathematical study of the interaction of a system of
connected elements (Berge, 1962; Köning, 1937). By investigating the
characteristics and interactions of predominant hashtags through the
principles of graph theory, we can widen our understanding of how social
network users perceive the CES provided by nature. \textbf{} We will
compare the outcomes of the application of graph theory network analysis
to image content analysis to assess the suitability and
cost-effectiveness of the methods. In addition, to attain a more
holistic assessment of CES provision, we will ascertain the diversity
and complementarity of the outcomes stemming from different social media
platforms.

The study focuses on three worldwide iconic coastal areas as
case-studies to illustrate the application of the proposed method,
namely the Great Barrier Reef (GBR) Marine Park in Australia, the
Galapagos Islands National Park in Ecuador and Easter Island National
Park in Chile. These areas include emblematic marine protected areas but
also protected terrestrial ecosystems. Two social media platforms,
Instagram and Twitter, were used as data sources. The high number of
users associated to these platforms and the markedly different content
format and user's needs and behaviours between platforms (Manikonda et
al., 2016) are expected to illustrate the diversity of CES stemming from
the case-study areas.{~}

The main objectives of the study were (i) to illustrate the application
of a novel methodology to assess the demand of CES through graph theory
network analysis, (ii) to compare the proposed methodology to existing
image content analysis techniques, and (iii) to explore the
complementarity of information extracted from different social media
platforms.{~}

To achieve these objectives, hashtag data from Instagram and Twitter
were analysed using graph theory analysis to identify emerging patterns
in CES demand. Additionally, manual and automatic identification of
Instagram image content was conducted for comparative purposes to assess
the alignment between both approaches. This exhaustive comparison of
remote assessment of CES allows insights into the most cost-efficient
techniques to undertake large-scale assessments of social perceptions on
ecosystems.{~}

\section{Methods}

\subsection{Data acquisition}

In June 2019, ten thousand posts were downloaded from Instagram for each
of the three case-study areas; similarly, ten thousand posts were
downloaded from Twitter per case study. After running a set of trials
with varying number of downloaded posts, the authors settled for 10,000
posts, as these were sufficient to capture the most frequently used
hashtags in each of the study areas. Instagram and Twitter posts were
downloaded through the corresponding application programming interface
(API). The Instagram public API is suitable for hashtag-based data
extraction, while Twitter required a developer account to access the
full history and volume of tweets. A specific development in R was made
by the authors for each API, both extracted similar data from social
media platforms. The API works as a keyword search method. For each case
study, a search query was executed, obtaining a set of 10,000 posts for
each area and platform that included the name of the area (i.e. query)
as a hashtag as part of the post.{~}

Relevant hashtags were used to extract the posts associated to the
case-studies: the hashtags ``\#greatbarrierreef'' and ``\#galapagos''
were used as queries for the GBR Marine Park and the Galapagos Islands
National Park respectively. While these hashtags represent the most
frequent way social media users refer to these areas in Instagram and
Twitter, based on the authors' observations, Easter Island was
frequently referred to as ``\#easterisland'', ``\#rapanui'' and
``\#isladepascua'', the last two representing the local names of the
area. Therefore, three separate posts' downloads were performed for
Easter Island using each of the three queries. Downloaded data for this
area were merged for subsequent analysis. Resulting datasets for each
case study were stored locally in a relational database.

Posts often contain non-relevant information as social media platforms
are frequently used as marketing and advertisement tools to reach a
wider public and often bots (automated data generating algorithms and
advertisements) are used to created large volumes of automated posts.
Datasets were filtered and cleaned in order to retain only relevant
information for further analysis (Di Minin et al., 2018; Varol et al.,
2017). Non-relevant hashtags, mostly related to advertisement, were
discarded.{~}

Since the main aim of the present study was to assess the type of CES
provided by the case-study areas, regardless of the user's nationality,
hashtags with a high frequency of appearance were translated to English
language. In addition, hashtags were scanned for spelling mistakes and
variations of the same word (e.g. bird - birds, traveller - traveler) in
order to standardise the dataset and avoid duplicates.{~}

\subsection{Image content analysis}

For each case study, photos associated to the original 10,000 posts were
also downloaded and stored for image content analysis. Two types of
analyses were performed, and their results compared: a manual procedure,
undertaken by the authors of this paper, and an automatic procedure
through machine learning technology.{~}

\subsubsection{Manual image content analysis}

The content of each image was visualised, analysed and classified using
an objective coding approach. To classify the CES in the case studies,
we adopted and modified the classification developed by Retka et al.
(2019) (table 1). In addition to the CES classification, we recorded
information on whether the photographs were taken above or below water,
on specific activities and on predominant habitat types and species
appearing on the photographs (Appendix 1).{~}

A random subsample of the photographs was drawn for each of the
case-studies for the image content assessment. To determine the minimum
number of photographs needed to assess the type of CES provided in each
case-study, cumulative frequency distributions were calculated and
plotted for each type of CES per case-study. Random sets of 10
photographs were assessed to quantify the presence of the different ES
classes. Additional sets of 10 photographs were sub-sampled and
classified until the cumulative average of the percentage of CES classes
stabilised.{~}

To assess the consistency of the classification criteria and the level
of agreement between the reviewers, a subsample of 75 random photographs
across the three case studies was evaluated by each reviewer. Cohen's
Kappa coefficient (Cohen, 1960) was used to assess the level of
agreement between the reviewers.

{[}table 1{]}

\subsubsection{Automated image content analysis}

The same set of photographs analysed manually was assessed through
Microsoft Captionbot Computer Vision's REST API
(https://azure.microsoft.com/en-gb/services/cognitive-services/computer-vision/).
CaptionBot is a free cognitive tool based on ComputerVision, a Microsoft
Azure cognitive service that distills relevant information from images.
CaptionBot does not require users to have experience in machine learning
but it provides powerful capabilities in content discovery, text
extraction and visual data processing to tag content from objects to
concepts, or extracting printed or handwritten text. Our intention was
to create an AI-based workflow using tools that were low cost but
equally adaptable and flexible. CaptionBot analyse Image method and
Python
({https://docs.microsoft.com/en-gb/azure/cognitive-services/Computer-vision/quickstarts/python-disk})
were used to obtain a JSON document containing the predictive response
with regards to the image content. The algorithm predictive response of
the content of the photographs was extracted in a natural language
format (e.g. ``I think it's a turtle swimming under water'', ``I'm not
sure but I think it's a man walking on the beach''). Based on the
information provided by Captionbot, the authors allocated each of the
photographs to one of the established CES classes. The level of
classification agreement between the manual and automatic classification
was assessed using Cohen's Kappa coefficient (Cohen, 1960).{~}

\subsection{Graph theory network analysis}

The analysis of networks using graph theory can be described as the
analysis of existing relationships between the different elements
contained in a network. The term \emph{vertex} is used to describe the
elements in a network, while the term \emph{edge} is used to refer to
the connections between the different vertices in a network. In our
case, vertices are represented by hashtags, while edges illustrate the
connections between hashtags (e.g. the hashtags included in the same
posts and the frequency of those connections).{~}

To assess relationships between hashtags and identify emerging
properties within the networks, we used centrality measures and
community structure detection algorithms. In networks consisting of
several vertices, some of them play a decisive role in facilitating a
large number of network connections. Such vertices are central in
network organization and are often identified by a range of metrics
known as centrality measures. Centrality measures are useful to
determine the relative importance of vertices and edges within the
overall network (Freeman, 1978). However, there are multiple
interpretations of what makes a vertex important and there are therefore
many measures of centrality (Freeman, 1978). Some commonly used measures
of centrality are: Degree; Betweenness; Closeness; Eigenvector
centrality; Kleinberg's hub centrality score (Hub score); Kleinberg's
authority centrality score (Authority score); and Page Rank.
Conceptually, the simplest form of centrality is \emph{Degree
centrality}, which represents the number of edges connected to a vertex.
In a social media network, where vertices are represented by hashtags,
the Degree centrality of a hashtag accounts for the number of
connections a hashtag has with other hashtags in the network. However,
not all connections are equally important. Connections with
well-connected vertices are more important than connections to vertices
that are poorly connected to others. Thus, a vertex is important if it
is connected to important neighbors, this is defined as Eigenvector
centrality. Therefore, it can happen that a vertex with high Degree
centrality has low Eigenvector values, e.g. a vertex could have many
links (i.e. high Degree) to poorly connected vertices (i.e. low
Eigenvector). Likewise, a vertex with few connections could have a high
Eigenvector centrality value if those few connections were to
well-connected vertices.{~}

In this study we focus on Eigenvector centrality measure to illustrate
social media data network structure. Eigenvector is a useful measure for
the analysis of hashtags in a social media network because it does not
necessarily highlight words with the highest frequency of occurrence
(e.g. \#instagram, \#instatravel, \#instaphoto, \#twitterpic), which
might not be necessarily informative. Eigenvector highlights hashtags
that are well connected with other hashtags related to the query search,
therefore, allows the emergence of relevant hashtags to understand the
structure of the network.{~}

In graph theory, a community is defined as a group of vertices where the
density of the edges between the vertices inside the group is greater
than the connections with the rest of the network. Vertices pertaining
to the same community display similar centrality measure values.
Generally, connections between vertices within the same community are
stronger than connections between vertices of different communities.
Here, in order to identify CES bundles, we organized the networks into
communities. Graphs depicting the social networks are composed of
vertices representing words.{~ }Word communities are the grammatical
contexts in which these words appear together. If the words are
mentioned frequently in the same context, they will form a community in
the graph. If they appear in different contexts, they will move away
from each other. To detect these communities, we applied the fast greedy
modularity optimization algorithm (Clauset et al., 2004).

Data mining, analyses and graphical outputs were generated using R, a
free and open source software (R Core Team, 2019). Specific R packages
were used to create hashtags networks, calculate centrality measures and
detect community structure (igraph, Csardi and Nepusz, 2006) and to
create network visualisations (ggraph v2.0., Pedersen, 2020). In the
community graphs, the order and distance of the communities to the
centre does not imply a greater degree of importance, it is a result of
the visualisation method.

\section{Results}

The focus of this study was to develop an innovative methodology for CES
assessment using social media and to compare it to existing
methodologies. For brevity and clarity, the results section focuses
particularly on one of the case-studies (GBR) to fully illustrate the
type of information obtained using graph theory network analysis, while
Galapagos and Easter Island are more succinctly explained (see Appendix
2 for figures).{~}

We first present results for the more direct and traditional methodology
of manual photo content analysis, move onto the automatic analysis of
photographs and finally report on the results obtained through our
proposed methodology. Results focus on ascertaining the type of CES
provided by each case study. {~}

\subsection{Image Analysis}

\subsubsection{Manual image content analysis}

A comparison of the 3 case studies revealed that the proportion of
underwater photographs in GBR (45\%) was markedly greater than in Easter
Island (11\%) or Galapagos (8\%). The predominant CES classes in GBR
were related to recreational activities (33\%), the appreciation of the
landscape and seascape (26\%) and nature (21\%), where the main subject
of the photographs was either fauna or flora. Snorkelling (14\%),
wildlife (14\%), diving (11\%), and habitat appreciation (7\%) comprised
the most popular activities depicted in the photographs (Figure 1). In
GBR approximately 40\% of the photographs coral reefs featured as the
main habitat and fish were the dominant animal group (17\%), featuring
either underwater or as recreational fishing trophies (Figure 1).

\begin{figure}
	\centering
	\includegraphics[width=1\linewidth]{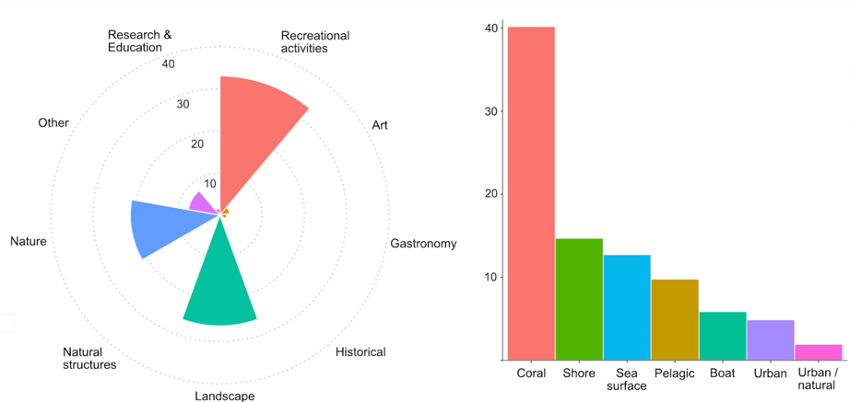}
	\caption{Instagram manual image content analysis for Great Barrier Reef. Left: percentage of photographs depicting specific CES. Right: percentage of photographs depicting specific habitats.}
	\label{fig:picture1}
\end{figure}

In Galapagos, predominant CES classes were nature (49\%), landscape
appreciation (19\%) and recreational activities (12\%). Nature
appreciation mainly focused on iconic wild animals: marine iguanas and
giant tortoises appeared on 21\% of the photographs, birds on 18\% and
marine mammals on 11\% of the photos (Appendix 2, Figure 1). Of the 3
case-studies, Galapagos was the area with the greatest proportion of
pictures focusing on wildlife (Galapagos 57\%, GBR 31\%, and Easter
Island 3\%). No particular habitat was frequently depicted, however 27\%
of photographs focused on the coastal shore fringe.

In Easter Island, a high proportion of photographs (38\%) were
classified within the historical monuments class (as Easter Island
statues featured frequently in photographs), followed by landscape
appreciation (13\%) and natural structures (11\%), such as volcano
craters and cliffs. Most photographs depicted grass fields (44\%) or
shorelines (16\%) (Appendix 2, Figure 2). Inter-reviewer Cohen's kappa
coefficient was high (0.87). {~}

\begin{tiny}
\begin{table}[]
	\caption{Cultural ecosystem service (CES) description used for the classification of photography content (Source: adapted from Retka et al., 2019)}
	\label{tab:table1}
	\begin{tabular}{p{5cm} p{6cm}}
		\hline
		\multicolumn{1}{c}{\textbf{CES Category}} & \multicolumn{1}{c}{\textbf{Description}}                                                                \\ \hline
		1. Artistic or cultural expressions and appreciation             & Photographs representing people in artistic activities or their products                                \\
		 \\
		2. Living cultural heritage                                      & Photographs representing people in cultural activities                                                  \\
		 \\
		 3. Gastronomy                                                    & Photographs representing typical meals/foods related to the area                                        \\
		 \\
		 4. Historical monuments                                          & Photographs depicting historical infrastructures (e.g. historical buildings, ruins)                     \\
		 \\
		 5. Landscape appreciation                                        & Photographs for which the main focus is a wide and large scale view of the landscape                    \\
		 \\
		 6. Nature appreciation                                           & Photographs focusing on fauna or flora                                                                  \\
		\\
		7. Natural structures and monuments                              & Photographs depicting a specific and well-defined landscape structure (e.g. cliff, cave)                \\
		 \\
		 8. Religious, spiritual or ceremonial activities & Photographs representing religious or spiritual monuments or activities (e.g. church, indigenous ritual) \\
		 \\
		 9. Research \& education                                         & Photographs showing research or education activities or equipment                                       \\
		 \\
		 10. Social recreation                                            & Photographs representing groups of people in an informal or non-dedicated recreative social environment \\
		 \\
		 11. Activity recreation                                          & Photographs showing people in a specific sports related activity                                        \\
		 \\
		 12. Other                                                        & Photographs that do not fit the above criteria                                                          \\ \hline
	\end{tabular}
\end{table}
\end{tiny}

\subsubsection{Automated image analysis}

The use of Microsoft Captionbot Computer Vision's REST API for the
automatic analysis of photograph content was deemed not satisfactory.
Cohen's Kappa coefficient values were low for all three case studies,
denoting that the level of agreement between the manual content analysis
performed by the authors and that of Captionbot was weak (GBR = 0.51,
Galapagos = 0.61, Easter Island = 0.40).{~}

Although overall CES class percentages were similar between the manual
and the automatic classification, the classification of individual
pictures was different, in GBR 61\% of the pictures were equally
classified, in Galapagos 72\% and in Easter Island 46\%.{~}

Captionbot capability of correctly describing photo content differed
between CES classes, as some classes were easier to capture than others.
While image content related to landscape, nature, recreational
activities or social interactions was identified by the algorithm, it
failed to detect CES classes related to research and education,
spirituality, art or historical monuments/heritage.{~}

\subsection{Network analysis}

\subsubsection{Great Barrier Reef}

In GBR, the Instagram graph visualization based on Eigenvector
centrality indicated that concepts related to underwater activities
(e.g. diving, snorkel, underwater photography hashtags), underwater life
(e.g. reef, coral, fish) and travel (e.g. travel, holidays) occupied
central positions within the network structure and were frequently
related to the ``greatbarrierreef'' hashtag (i.e. the query). These
hashtags had high Eigenvector values, indicating that they frequently
appeared on GBR related posts and at the same time were related to
concepts also appearing frequently. In addition, high Eigenvector values
revealed geographical locations frequently related to popular hashtags,
e.g. ``whitsundays'' was well connected to ``nature''. Concepts related
to positive and ``feel-good'' aspects (e.g. love, happiness, beach life,
fun) were often located surrounding the core concepts on the centre
although did not occupy central positions. Hashtags related to
environmental awareness also featured as part of the network (e.g.
global warming) but did not occupy a central position in the structure
(Figure 2).

\begin{figure}
	\centering
	\includegraphics[width=1\linewidth]{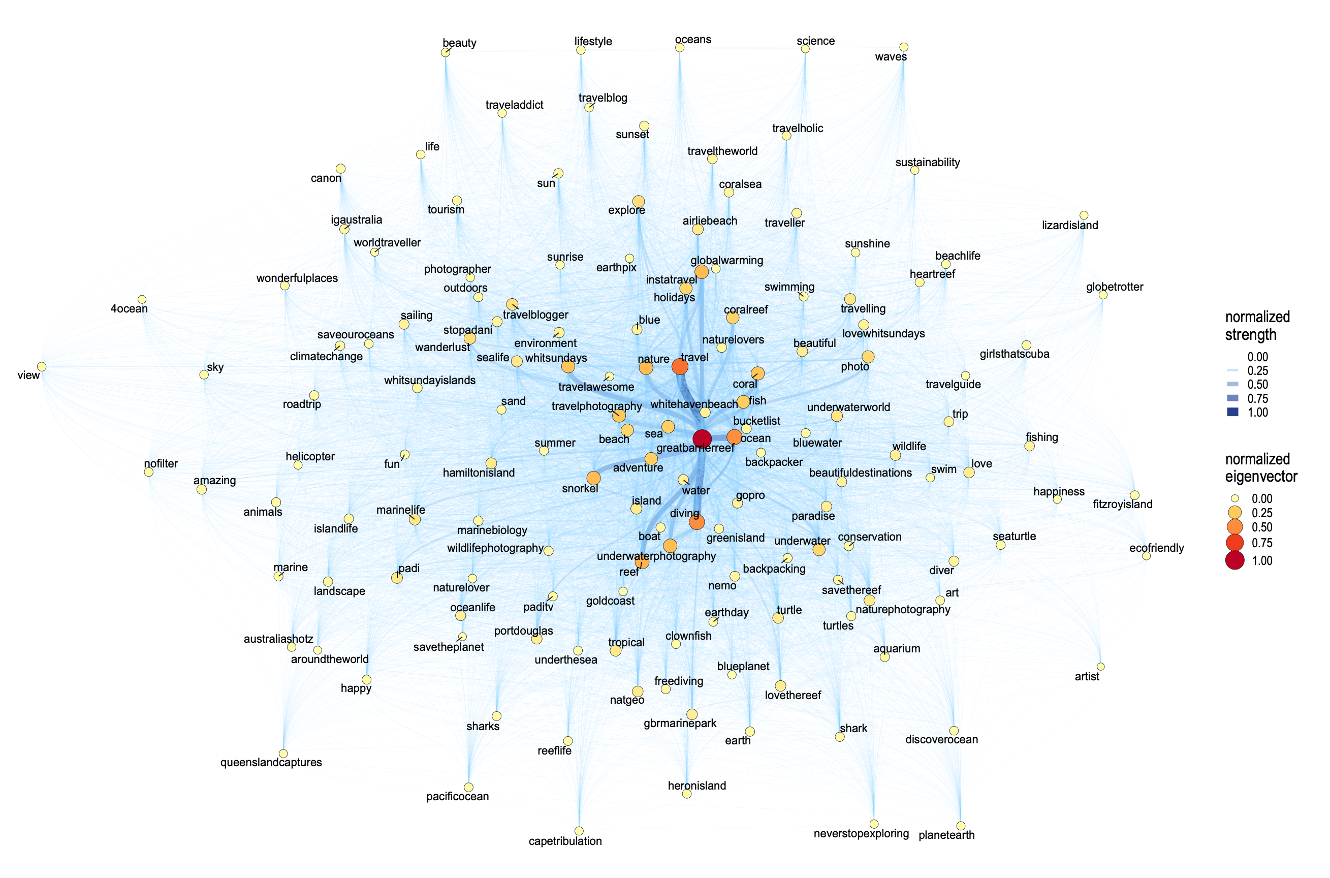}
	\caption{Great Barrier Reef Instagram Eigenvector network}
	\label{fig:picture2}
\end{figure}

In GBR, the community detection algorithm grouped the different hashtags
into 3 overarching themes (Figure 3). Hashtags in the first community
were mainly related to the underwater marine world (e.g. reef, ocean
life, shark) and associated recreational activities (e.g., free diving,
snorkel, underwater photography), as well as to concepts related to
environmental conservation (e.g. sustainability, conservation, earth
day). This community identified key habitats and species as providers of
CES, such as coral, fish, turtles or sharks.{~}

\begin{figure}
	\centering
	\includegraphics[width=1\linewidth]{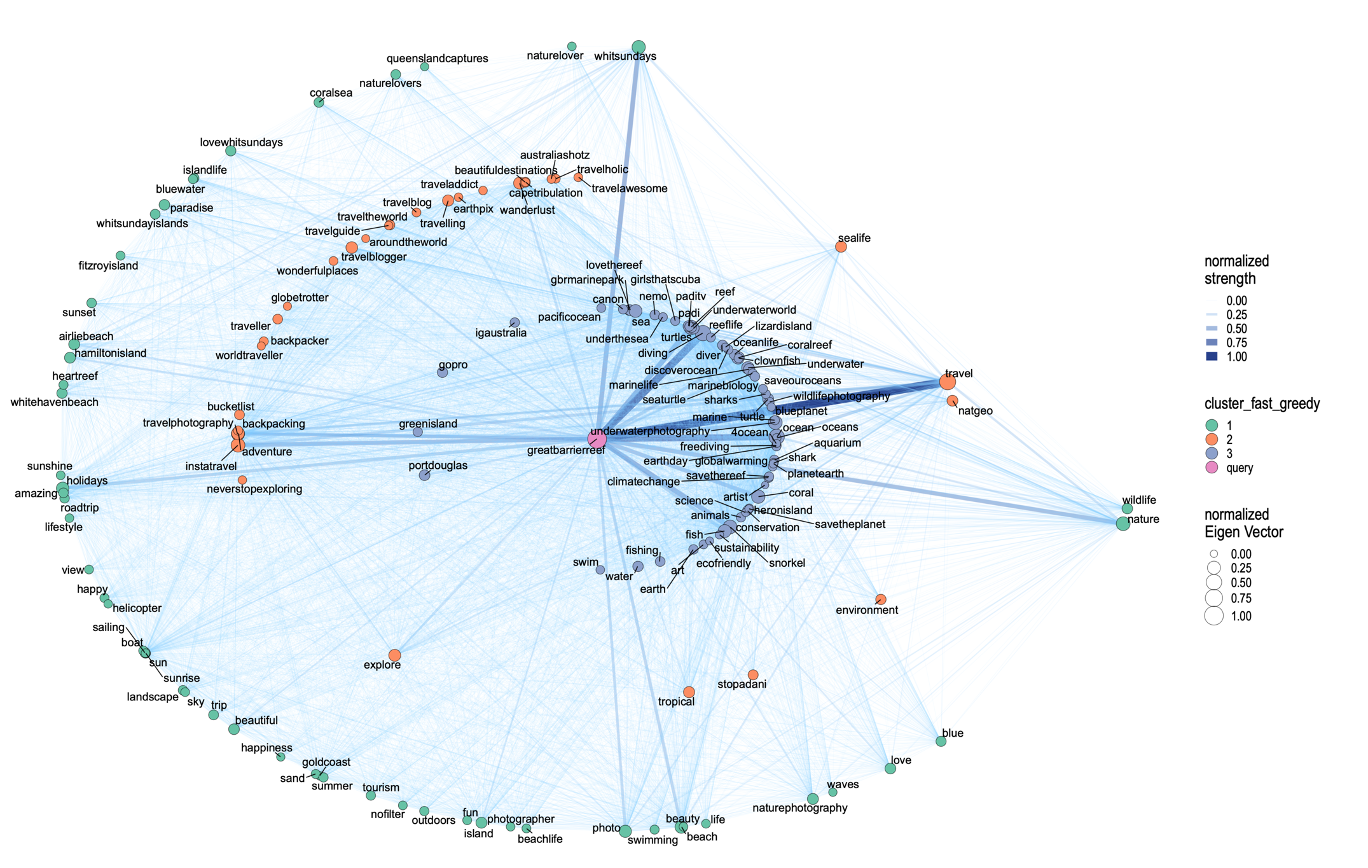}
	\caption{Hashtags communities generated through Fast Greedy algorithm from Great Barrier Reef Instagram Eigenvector network}
	\label{fig:picture3}
\end{figure}

In addition, cognitive services also featured in this community (e.g.
science, marine biology) with lower eigenvector values. The second
community was predominantly dominated by hashtags related to travelling
(e.g. travel, wanderlust, travelholic) and what travelling allows, such
as fulfilling life long wishes (e.g. bucket list), reaching remote
places (e.g. wonderful places, around the world, beautiful destinations)
or creating feelings of adventure (e.g. explore, adventure, never stop
exploring). In the third community, hashtags with greater Eigenvector
values were related to the provision of nature and wildlife holidays,
the feelings those experiences create (e.g. fun, happiness, love),
activities enjoyed (e.g. swimming, outdoors, boat, sailing), memorable
moments and places and descriptions associated to them (e.g. sunset,
sunrise, beauty, amazing).

In the Eigenvector-based Twitter visualization, hashtags with greater
Eigenvector values and most frequently connected to the query
(``greatbarrierreef''), mainly revolved around climate change (e.g.
climate action, climate crisis, climate emergency) and the environment
(e.g. coral, ocean, reef, nature), creating a strong environmental
awareness theme (Figure 4).{~}

\begin{figure}
	\centering
	\includegraphics[width=1\linewidth]{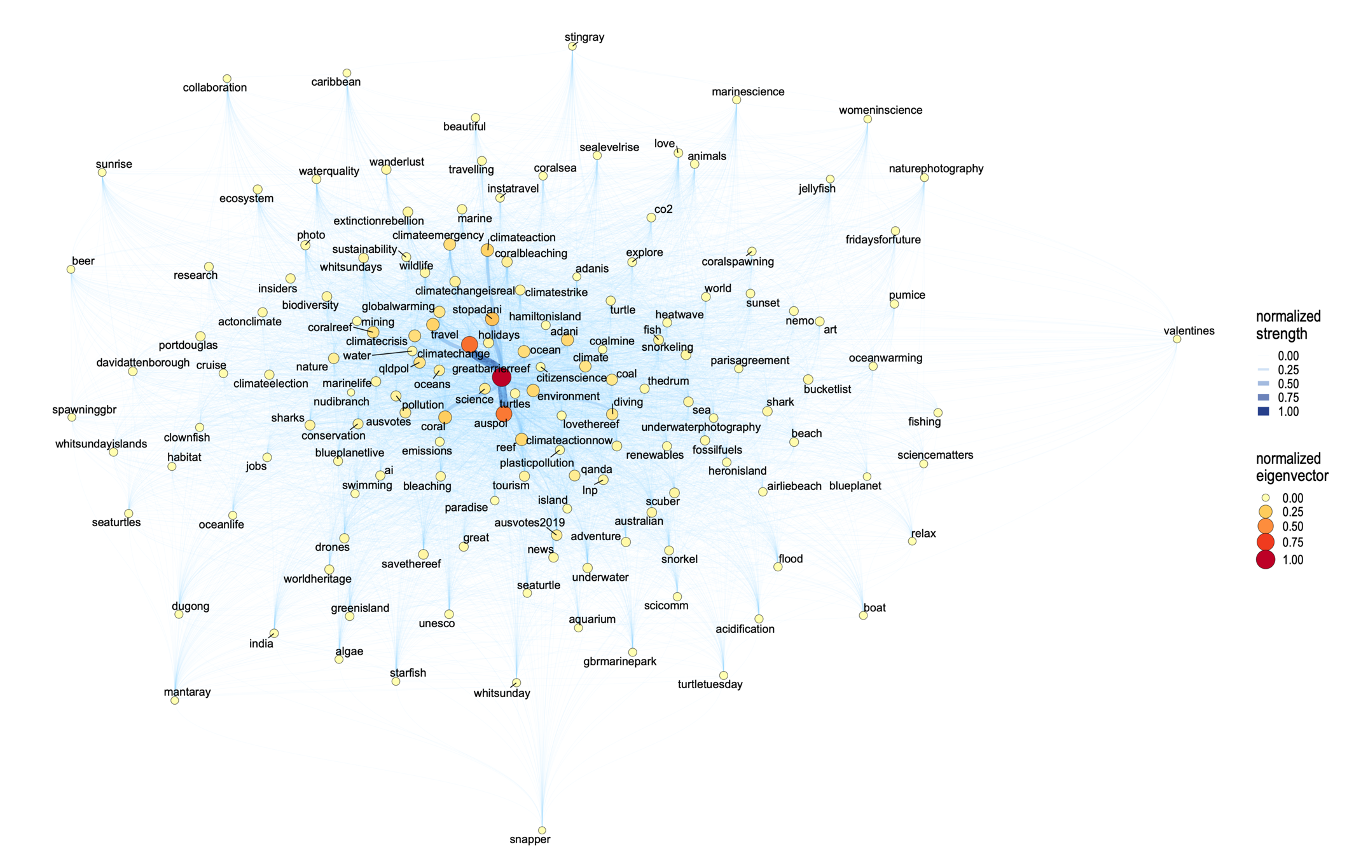}
	\caption{Great Barrier Reef Twitter Eigenvector network}
		\label{fig:picture4}
\end{figure}

Hashtags related to underwater activities and specific underwater life
were present in the network, although they were mostly located towards
the periphery of the network, displaying a more secondary role.

Three communities were detected in the Twitter GBR network (Figure
5).{~}

\begin{figure}
	\centering
	\includegraphics[width=1\linewidth]{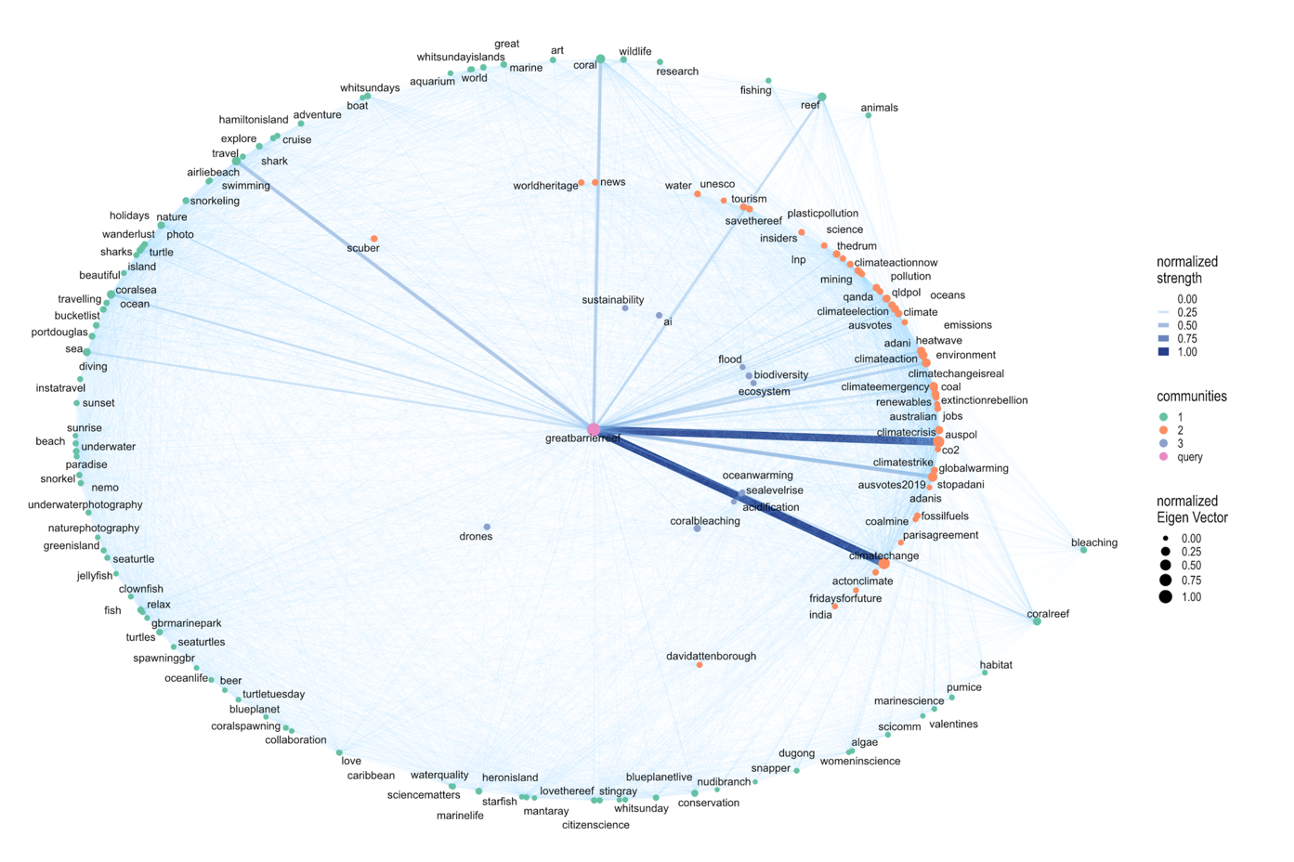}
	\caption{Hashtags communities generated through Fast Greedy algorithm from Great Barrier Reef Twitter Eigenvector network}
	\label{fig:picture5}
\end{figure}

The community closest to the centre was dominated by marine
environmental science concepts (e.g. bleaching, sea level rise, ocean
warming, acidification) and marine life (e.g. shark, reef, algae). The
second community followed a political discourse around climate change
and action for change and sustainability (e.g. Fridays for future,
climate strike, renewables, extinction rebellion). The third community
rotated around holidays, travelling and memorable moments (e.g. beach,
paradise, sunset), recreational activities (e.g. boat, scuba,
snorkelling) and marine life (e.g. snapper, nudibranch, starfish).{~}

\subsubsection{Galapagos}

In Galapagos' Eigenvector-based Instagram visualization, the vertices
with greatest Eigenvector values were related to nature and wildlife
concepts in general and to specific animal groups in particular (e.g.
sea lion, birds, iguana, tortoise). Travel, photography and diving were
also prominent hashtags within the network. However, overall, the
Galapagos network was mostly dominated by wildlife related hashtags
(Figure 3, Appendix 2).{~}

In Galapagos Instagram network, hashtags were grouped into five
communities (Figure 4, Appendix 2). The first community was dominated by
hashtags related to the marine environment, underwater marine life and
associated recreational activities. Hashtags in the second community
revolved around travel and the desire for travelling, in a similar way
as in GBR. The third community also centred on travelling but from an
adventurous and laidback approach. The fourth community was dominated by
wildlife and nature aspects, as well as by the love for nature and
conservation values. A high number of hashtags representing different
animal groups were allocated to this community. The fifth community
related to life and nice feelings. {~}

In Galapagos' Eigenvector-based Twitter visualization (Figure 5,
Appendix 2), hashtags with greater Eigenvector values were similar to
those in Instagram. Nature, wildlife and travel concepts occupied the
most central positions in the network and had greater Eigenvector
values. Hashtags referring to iconic animals were also centrally
positioned.{~}

Hashtags were clustered into two big groups by the community detection
algorithm (Figure 6, Appendix 2). In the community closer to the centre,
all hashtags had similar eigenvector values and turned around iconic
wildlife groups and recreational activities. The second community
revolved around travelling to locations that enable the enjoyment of
wildlife and nature.

\subsubsection{Easter Island}

Instagram Easter Island's network was dominated by concepts related to
travelling, culture and cultural identity, which occupied central
positions in the Eigenvector-based visualisation (Figure 7, Appendix 2).
Photography and aesthetics concepts also featured frequently although
they were not central.{~}

Hashtags were grouped into four communities (Figure 8, Appendix 2). The
central community was explicitly related to underwater recreational
activities, in particular diving and underwater photography. The second
community mostly revolved around the cultural heritage of Easter Island
as it included hashtags such as music, sculpture, architecture or
archaeology. The third community was dominated by travel related
hashtags. The last community was more heterogeneous as it bundled
concepts of cultural identity, nature and ``living a good life''
concepts.{~}

In Twitter, aside from the hashtags used to build the network, the
hashtag ``chile'' (country where the case-study is located) and ``moai''
(the monolithic human figures carved by the Rapanui people), the rest of
the hashtags had low Eigenvector values (Figure 9, Appendix 2). Hashtags
related to travel, aesthetics, heritage and holidays were located in
central positions; however, their associated Eigenvector values were
low, and no particular trend was noticeable. Concepts related to aspects
related to living a meaningful life were located towards the periphery
of the network.

The resulting communities from Easter Island Twitter's network presented
a more miscellaneous typology and were not as defined as the ones in the
other two case-studies (Figure 10, Appendix 2). Hashtags were grouped
into three communities. Starting from the centre, the first community
revolved mainly around historical and cultural heritage. No discernible
overall discourse was detected in the second community, as it bundled
hashtags related to travel, photography and nature, among others. The
majority of hashtags in the third cluster were in Spanish language and
no obvious pattern emerged from this cluster.{~}

\section{Discussion}

The focus of this study was to introduce a novel methodology for CES
assessment using social media and to compare it to existing methods, and
to validate an efficient methodology for a more encompassing local and
global CES assessments in marine and coastal areas. The analysis of
social media data from two different platforms and the application of
different methodological approaches allowed us to establish comparisons
in terms of the outputs obtained and the cost-effectiveness of the
methods used.{~}

In summary, the use of graph theory to analyse social media data
provided a more holistic perspective on the assessment of CES, including
a range of values, from tangible recreational activities to intangible
values related to feelings and perceptions. Manual image content
analysis provided a thorough assessment of uses, values and ecosystem
preferences, but failed short to capture intangible aspects such as
relational values.

As in previous studies, we ascertained how the manual image content
analysis of social media photographs provides in-depth information on
the CES classes (e.g., Oteros-Rozas et al., 2018), activities (e.g.,
Wood et al., 2013) and benefits (e.g., Gliozzo et al., 2016) arising
from specific areas, as well as on the habitat types and species{~
}providing those benefits (e.g., Sbragaglia et al., 2019). However, the
application of this method is extremely time consuming (approximately,
2-4 min per picture) and while it might be a suitable approach for small
geographical scale applications (e.g., Clemente et al., 2019; Hausmann
et al., 2017; Retka et al., 2019; Schirpke et al., 2018), it is not a
cost-effective methodology for large scale assessments.{~}

The automatic alternative for photo content identification through the
application of Captionbot automated image recognition using natural
language outputs was considered not satisfactory, as the level of
agreement between human and machine-based CES classification was too
low. That is not to say that the application of machine learning
techniques for automated image recognition in general is not suitable
for CES assessments, as there is an increasing number of studies that
have successfully applied it in terrestrial areas (e.g.,Lee et al.,
2019; Oteros-Rozas et al., 2018). Several factors might have contributed
to the low quality of our results. As a rule of thumb, the higher the
number of pictures used in the training of machine learning algorithms,
the more accurate the predictive results are (Mikołajczyk and
Grochowski, 2018). Therefore, accurate results are most frequent for
those environments for which considerable volumes of photographs are
available. In general, while terrestrial pictures are overly abundant,
pictures related to underwater environments are fewer in comparison,
mainly due to the more restrictive nature of this environment, therefore
limiting the accuracy levels available for those contexts. A second
factor influencing the quality of the results is the automatic
classification algorithm used, as it will determine the finesse of the
results obtained (Martin-Abadal et al., 2020). The level and types of
CES classes detected through machine learning can range from coarse
levels of CES classifications (Richards et al., 2018), that might fall
short in terms of management applicability, to more holistic
classifications including the detection of existence values (Lee et al.,
2019). Finally, a different automated recognition approach might have
contributed to the achievement of better results in our study. As an
alternative to focusing on the single output in the form of natural
language for each of the photographs used here, the automatic
assignation of multiple tags for each picture and subsequent analysis of
the tags might have generated different results (e.g., Lee et al.,
2019). However, the greatest challenge in the use of automatic picture
classification for CES assessment is that algorithms are not yet capable
of capturing intangible aspects such as spiritual or cultural heritage
values. More important perhaps, is the fact that photographs cannot
always convey elusive aspects such as feelings, social ties or cognitive
values (Lee et al., 2019). Comparatively, we have shown how the use of
graph theory network analysis on user's text has the capacity of
encompassing those aspects.{~}

The application of graph theory network analysis revealed the most
frequent concepts arising from each of the case study areas and the
interactions between them. This approach facilitated the identification
of the most popular hashtags, provided information on which popular
hashtags were related to other popular hashtags and how the different
hashtags grouped together. It provided information on three different
fronts; (i) what the main CES stemming from the area were and how
different services tended to appear together as CES bundles; (ii) on ES
providers (ESP), that is, the main elements, including habitats, species
or natural structures supporting CES; and (iii) finally, on the
frequency of linkages between geographical hashtags and activities or
benefits hashtags allowing the extraction of information regarding
popular places where there is demand for particular CES. Previous social
media image analysis studies have also demonstrated the capacity to
identify CES bundles (Oteros-Rozas et al., 2018), ESP providers (Arbieu
et al., 2018; Hausmann et al., 2017) and the spatial distribution of CES
provision and demand (Clemente et al., 2019; Fischer et al., 2018; Gosal
et al., 2019). However, while the analysis of photographs mostly offers
a vision of ``what the eye can see'', the analysis of associated text
offers a perspective beyond the material or instrumental values
associated to nature to move into the realm of relational values.{~}

The concept of relational values, introduced by Chan et al. (2016),
broadens the notions of intrinsic and instrumental values to include the
values relative to the meaningfulness of relationships between people
and nature (Stenseke, 2018). The hope is that the adoption of a
relational value framework will be more inclusive of known aspects of
wellbeing, such as connections to others and to nature or place
attachment and identity (Chan et al., 2016). Along these lines, concepts
or relations related to nature-inclusive eudaimonia are also a form of
relational values (Knippenberg et al., 2018). Eudaimonic values can be
defined as the values associated with living a good and meaningful life
(Ryan and Deci, 2001; Ryff and Singer, 2008). ``Nature-inclusive
eudaimonia'', as introduced by Knippenberg (2018), goes further and
encompasses a flourishing life in which nature is an integral part of
wellbeing. Here, we show how the analysis of hashtags through graph
theory network offers a description made by the user itself that often
includes relational values aspects.

Previous studies based on social media image analysis have already made
some first steps towards capturing values pertaining to the realm of
relational values, such as existence (Martínez Pastur et al., 2016),
spiritual or social values and relations (Oteros-Rozas et al., 2018).
However, these studies have been based on proxies, such as the presence
of places serving as meeting points with friends being equal to social
values (Oteros-Rozas et al., 2018), or the assumption that pictures
focusing on nature appreciation are equivalent to an existence value
(Martínez Pastur et al., 2016). Although these assumptions and
associations might be sound, they are inherently linked to a
researcher's interpretation bias, therefore perhaps not fully or rightly
capturing the meaning of the picture. Conversely, the analysis of users'
posts allows to capture eudaimonic notions of living a good life through
sharing experiences with those who are more important to us (family,
friends\ldots), positive feelings emerging from being in contact with
and surrounded by nature (e.g., happiness, fun, love), the love for
nature (e.g., nature lovers, love the reef), the urge and need to
preserve nature (e.g., save our oceans, save the planet) or cultural
identity aspects (e.g., tapati, chilepo) in descriptions made by the
users, minimizing thus interpreter's bias. We, therefore, conclude that
the use of graph theory network analysis on social media hashtags offers
a different level of nuanced comprehension of relational value aspects
when compared to the analysis of photo content. It offers a window
through which we can contemplate the different relational aspects that
people experience when in contact with nature, minimising the potential
distortions associated to interpreter's bias.{~}

Despite the potential of social media data, our results show that the
consideration of outputs stemming from different platforms is an
important aspect to consider. Although the themes emerging from
Instagram and Twitter were similar, their magnitude of centrality, and
therefore importance, differed within the case studies analysed. While
Twitter generally reflected what users' thought, including their
political and environmental views and concerns, Instagram contained
information at a more emotional and relational level, as it focused on
what people do and want to show. These aspects have also been captured
by other studies, as Manikonda et al. (2016) describes it through the
expression ``tweeting the mind and instagramming the heart'' (Manikonda
et al., 2016). Therefore, the selection of social media platform will be
conditioned by the goals of the study, as well as target population
group. Nevertheless, we argue that data integration offers a more
comprehensive understanding of the different values held be people on
nature. It is also important to highlight that in general, studies based
on social media data will be restricted by the volume of posts available
for the areas under study.{~}

Generally, CES assessment studies using social media data make use of
geo-located photographs. This allows for the identification of spatial
distribution patterns of CES demand (e.g., Clemente et al., 2019), which
is relevant in terms of natural areas conservation and management. Due
to Instagram and Twitter privacy policy, the download of geo-tagged data
was not possible in our study. However, although not as precise as the
use of geo-tagged photographs, the recurrent presence of geographical or
places hashtags, frequently appearing together with certain activities
(e.g. diving, hiking) also allows for the identification of hotspots of
activity. Despite this shortcoming, we conclude that the application of
graph theory network analysis on social media data can be considered a
cost-effective method due to the short time needed to process the data
and its applicability at multiple spatial scales, as it can be used at
local to global scales through the simultaneous analysis of different
locations.{~}

Emerging patterns have the potential to be useful for managers and
policy makers to (i) identify and establish relationships between
subjective CES perceptions in the area; (ii) identify potential
trade-offs between different CES and between CES and other ES; (iii) and
to identify policy measures needed to preserve CES.

\section{Conclusions}

Social media data has emerged as a powerful source of information to
assess the indirect provision of CES. Generally, studies focus on the
analysis of data stemming from single platforms. In addition, the most
widely used method is based on the analysis of photo content, which
offers a partial vision of the range of CES offered by nature. Partial
in terms of the type of captured values and associated interpreter's
bias. Here, we introduce graph theory network analysis as a novel way to
analyse different sources of social media data to asses CES. We conclude
that the analysis of hashtags associated to social media posts using
graph theory offer information, not only on the instrumental values
associated to nature, but go further and provide useful information on
human-nature relational aspects and eudaimonic concepts. These are
aspects that photo content analysis has not yet been able to fully
capture. We also highlight the importance of considering data from
different social media platforms as the type of users and information
offered by the different platforms highlight different CES aspects.
Resulting networks are a reflection of the interactions between the
social media platform used and the environmental and cultural
characteristics of the area under consideration. As an example, in
Instagram, GBR users tend to share their coral reef diving experiences,
highlighting aspects of adventure and discovery. While in Easter Island,
Twitter highlights aspects related to cultural heritage preservation.
Thus, the combination of the social media platform and the cultural and
environmental characteristics of the area, establish a framework of
content possibilities from which the users tend to highlight certain
aspects. The ease of application and short computing processing times
involved in the retrieval and analysis of the data makes the use of
graph theory network a cost-effective method with the potential of being
applied to large geographical scales.{~}

\section*{Acknowledgements}

This work is a result of the ECOMAR Network, ``Evaluation and monitoring
of marine ecosystem services in Iberoamérica'' (project number
417RT0528) funded by the CYTED program. During the time of the study and
writing period ARF was supported by a H2020-Marie Skłodowska-Curie
Action MSCA-IF-2014 (ref. 655475); AOA was supported by a H2020-Marie
Skłodowska-Curie Action MSCA-IF-2016 (ref. 746361). SdJ was supported by
a H2020-Marie Skłodowska-Curie Action MSCA-IF-2016 (ref. 743545). PP was
funded by the Xunta de Galicia (RECREGES II Project, Grant ED481B2018/
017).

\section*{References}

\begin{scriptsize}
Arbieu, U., Grünewald, C., Martín-López, B., Schleuning, M.,
Böhning-Gaese, K., 2018. Large mammal diversity matters for wildlife
tourism in Southern African Protected Areas: Insights for management.
Ecosyst. Serv. 31, 481--490.
https://doi.org/10.1016/j.ecoser.2017.11.006

Berge, C., 1962. The theory of graphs and its applications. Bull. Math.
Biophys. 24, 441--443. https://doi.org/10.1007/bf02478000

Bratman, G.N., Anderson, C.B., Berman, M.G., Cochran, B., de Vries, S.,
Flanders, J., Folke, C., Frumkin, H., Gross, J.J., Hartig, T., Kahn,
P.H., Kuo, M., Lawler, J.J., Levin, P.S., Lindahl, T., Meyer-Lindenberg,
A., Mitchell, R., Ouyang, Z., Roe, J., Scarlett, L., Smith, J.R., van
den Bosch, M., Wheeler, B.W., White, M.P., Zheng, H., Daily, G.C., 2019.
Nature and mental health: An ecosystem service perspective. Sci. Adv. 5,
eaax0903. https://doi.org/10.1126/sciadv.aax0903

Chan, K.M.A., Balvanera, P., Benessaiah, K., Chapman, M., Díaz, S.,
Gómez-Baggethun, E., Gould, R., Hannahs, N., Jax, K., Klain, S., Luck,
G.W., Martín-López, B., Muraca, B., Norton, B., Ott, K., Pascual, U.,
Satterfield, T., Tadaki, M., Taggart, J., Turner, N., 2016. Why protect
nature? Rethinking values and the environment. Proc. Natl. Acad. Sci. U.
S. A. 113, 1462--1465. https://doi.org/10.1073/pnas.1525002113

Chan, K.M.A., Guerry, A.D., Balvanera, P., Klain, S., Satterfield, T.,
Basurto, X., Bostrom, A., Chuenpagdee, R., Gould, R., Halpern, B.S.,
Hannahs, N., Levine, J., Norton, B., Ruckelshaus, M., Russell, R., Tam,
J., Woodside, U., 2012. Where are Cultural and Social in Ecosystem
Services? A Framework for Constructive Engagement. Bioscience 62,
744--756. https://doi.org/10.1525/bio.2012.62.8.7

Clauset, A., Newman, M.E.J., Moore, C., 2004. Finding community
structure in very large networks. Phys. Rev. E - Stat. Physics, Plasmas,
Fluids, Relat. Interdiscip. Top. 70, 6.
https://doi.org/10.1103/PhysRevE.70.066111

Clemente, P., Calvache, M., Antunes, P., Santos, R., Cerdeira, J.O.,
Martins, M.J., 2019. Combining social media photographs and species
distribution models to map cultural ecosystem services: The case of a
Natural Park in Portugal. Ecol. Indic. 96, 59--68.
https://doi.org/10.1016/j.ecolind.2018.08.043

Cohen, J., 1960. A Coefficient of Agreement for Nominal Scales. Educ.
Psychol. Meas. 20, 37--46. https://doi.org/10.1177/001316446002000104

Csardi, G., Nepusz, T., 2006. The igraph software package for complex
network research. Interjournal Complex Sy, 1695.

Di Minin, E., Fink, C., Tenkanen, H., Hiippala, T., 2018. Machine
learning for tracking illegal wildlife trade on social media. Nat. Ecol.
Evol. https://doi.org/10.1038/s41559-018-0466-x

Di Minin, E., Tenkanen, H., Toivonen, T., 2015. Prospects and challenges
for social media data in conservation science. Front. Environ. Sci. 3,
63. https://doi.org/10.3389/fenvs.2015.00063

Erlandson, J.M., Rick, T.C., 2010. Archaeology Meets Marine Ecology: The
Antiquity of Maritime Cultures and Human Impacts on Marine Fisheries and
Ecosystems. Ann. Rev. Mar. Sci. 2, 231--251.
https://doi.org/10.1146/annurev.marine.010908.163749

Everard, M., Jones, L., Watts, B., 2010. Have we neglected the societal
importance of sand dunes? An ecosystem services perspective. Aquat.
Conserv. Mar. Freshw. Ecosyst. 20, 476--487.
https://doi.org/10.1002/aqc.1114

Fischer, L.K., Honold, J., Botzat, A., Brinkmeyer, D., Cvejić, R.,
Delshammar, T., Elands, B., Haase, D., Kabisch, N., Karle, S.J.,
Lafortezza, R., Nastran, M., Nielsen, A.B., van der Jagt, A.P.,
Vierikko, K., Kowarik, I., 2018. Recreational ecosystem services in
European cities: Sociocultural and geographical contexts matter for park
use. Ecosyst. Serv. 31, 455--467.
https://doi.org/10.1016/j.ecoser.2018.01.015

Freeman, L.C., 1978. Centrality in social networks conceptual
clarification. Soc. Networks 1, 215--239.
https://doi.org/10.1016/0378-8733(78)90021-7

Garcia Rodrigues, J., Conides, A., Rivero Rodriguez, S., Raicevich, S.,
Pita, P., Kleisner, K., Pita, C., Lopes, P., Alonso Roldán, V., Ramos,
S., Klaoudatos, D., Outeiro, L., Armstrong, C., Teneva, L., Stefanski,
S., Böhnke-Henrichs, A., Kruse, M., Lillebø, A., Bennett, E., Belgrano,
A., Murillas, A., Sousa Pinto, I., Burkhard, B., Villasante, S., 2017.
Marine and Coastal Cultural Ecosystem Services: knowledge gaps and
research priorities. One Ecosyst. 2.
https://doi.org/10.3897/oneeco.2.e12290

Ghermandi, A., Sinclair, M., 2019. Passive crowdsourcing of social media
in environmental research: A systematic map. Glob. Environ. Chang. 55,
36--47. https://doi.org/10.1016/j.gloenvcha.2019.02.003

Gliozzo, G., Pettorelli, N., Muki Haklay, M., 2016. Using crowdsourced
imagery to detect cultural ecosystem services: A case study in South
Wales, UK. Ecol. Soc. 21. https://doi.org/10.5751/ES-08436-210306

Gosal, A.S., Geijzendorffer, I.R., Václavík, T., Poulin, B., Ziv, G.,
2019. Using social media, machine learning and natural language
processing to map multiple recreational beneficiaries. Ecosyst. Serv.
38, 100958. https://doi.org/10.1016/J.ECOSER.2019.100958

Hausmann, A., Toivonen, T., Slotow, R., Tenkanen, H., Moilanen, A.,
Heikinheimo, V., Di Minin, E., 2017. Social Media Data Can Be Used to
Understand Tourists' Preferences for Nature-Based Experiences in
Protected Areas. Conserv. Lett. https://doi.org/10.1111/conl.12343

IPBES, 2019. Summary for policymakers of the regional assessment report
on biodiversity and ecosystem services for Asia and the Pacific of the
Intergovernmental Science-Policy Platform on Biodiversity and Ecosystem
Services, IPBES.

Klain, S.C., Satterfield, T.A., Chan, K.M.A., 2014. What matters and
why? Ecosystem services and their bundled qualities. Ecol. Econ. 107,
310--320. https://doi.org/10.1016/j.ecolecon.2014.09.003

Knippenberg, L., de Groot, W.T., van den Born, R.J., Knights, P.,
Muraca, B., 2018. Relational value, partnership, eudaimonia: a review.
Curr. Opin. Environ. Sustain. 35, 39--45.
https://doi.org/10.1016/j.cosust.2018.10.022

Köning, D., 1937. Theorie der endlichen und unendlichen Graphen.
Monatshefte für Math. und Phys. 46, A17--A18.
https://doi.org/10.1007/bf01792729

Lee, H., Seo, B., Koellner, T., Lautenbach, S., 2019. Mapping cultural
ecosystem services 2.0 -- Potential and shortcomings from unlabeled
crowd sourced images. Ecol. Indic. 96, 505--515.
https://doi.org/10.1016/j.ecolind.2018.08.035

Leenhardt, P., Low, N., Pascal, N., Micheli, F., Claudet, J., 2015. The
role of marine protected areas in providing ecosystem services, Aquatic
Functional Biodiversity: An Ecological and Evolutionary Perspective.
https://doi.org/10.1016/B978-0-12-417015-5.00009-8

Mancini, F., Coghill, G.M., Lusseau, D., 2018. Using social media to
quantify spatial and temporal dynamics of nature-based recreational
activities. PLoS One 13. https://doi.org/10.1371/journal.pone.0200565

Manikonda, L., Meduri, V.V., Kambhampati, S., 2016. Tweeting the Mind
and Instagramming the Heart: Exploring Differentiated Content Sharing on
Social Media, in: Proceedings of the Tenth International AAAI Conference
on Web and Social Media (ICWSM 2016). pp. 639--642.

Martin-Abadal, M., Ruiz-Frau, A., Hinz, H., Gonzalez-Cid, Y., 2020.
Jellytoring: Real-Time Jellyfish Monitoring Based on Deep Learning
Object Detection. Sensors 2020, Vol. 20, Page 1708 20, 1708.
https://doi.org/10.3390/S20061708

Martínez Pastur, G., Peri, P.L., Lencinas, M. V., García-Llorente, M.,
Martín-López, B., 2016. Spatial patterns of cultural ecosystem services
provision in Southern Patagonia. Landsc. Ecol. 31, 383--399.
https://doi.org/10.1007/s10980-015-0254-9

McMichael, A., Scholes, R., Hefny, M., Pereira, E., Palm, C., Foale, S.,
2005. Linking ecosystem services and human well-being, in: Norgaard, R.,
Wilbanks, T. (Eds.), Ecosystems and Human Well-Being\,: Multi-Scale
Assessments. Millenium Ecosystem Assessment Series, 4. Island Press,
Washington, DC, USA, pp. 43--60. https://doi.org/30

Mikołajczyk, A., Grochowski, M., 2018. Data augmentation for improving
deep learning in image classification problem, in: 2018 International
Interdisciplinary PhD Workshop, IIPhDW 2018. Institute of Electrical and
Electronics Engineers Inc., pp. 117--122.
https://doi.org/10.1109/IIPHDW.2018.8388338

Ortiz-Ospina, E., 2019. The rise of social media - Our World in Data
{[}WWW Document{]}. URL https://ourworldindata.org/rise-of-social-media
(accessed 3.21.20).

Oteros-Rozas, E., Martín-López, B., Fagerholm, N., Bieling, C.,
Plieninger, T., 2018. Using social media photos to explore the relation
between cultural ecosystem services and landscape features across five
European sites. Ecol. Indic. 94, 74--86.
https://doi.org/10.1016/j.ecolind.2017.02.009

Pedersen, T.L., 2020. ggraph: An Implementation of Grammar of Graphics
for Graphs and Networks.

PRC, 2019. Use of different online platforms by demographic groups.

R Core Team, 2019. R: A language and environment for statistical
computing.

Retka, J., Jepson, P., Ladle, R.J., Malhado, A.C.M., Vieira, F.A.S.,
Normande, I.C., Souza, C.N., Bragagnolo, C., Correia, R.A., 2019.
Assessing cultural ecosystem services of a large marine protected area
through social media photographs. Ocean Coast. Manag. 176, 40--48.
https://doi.org/10.1016/j.ocecoaman.2019.04.018

Richards, D.R., Friess, D.A., 2015. A rapid indicator of cultural
ecosystem service usage at a fine spatial scale: Content analysis of
social media photographs. Ecol. Indic. 53, 187--195.
https://doi.org/10.1016/j.ecolind.2015.01.034

Richards, D.R., Tunçer, B., Tunçer, B., 2018. Using image recognition to
automate assessment of cultural ecosystem services from social media
photographs. Ecosyst. Serv. 31, 318--325.
https://doi.org/10.1016/j.ecoser.2017.09.004

Ryan, R.M., Deci, E.L., 2001. On Happiness and Human Potentials: A
Review of Research on Hedonic and Eudaimonic Well-Being. Annu. Rev.
Psychol. 52, 141--166. https://doi.org/10.1146/annurev.psych.52.1.141

Ryff, C.D., Singer, B.H., 2008. Know thyself and become what you are: A
eudaimonic approach to psychological well-being. J. Happiness Stud. 9,
13--39. https://doi.org/10.1007/s10902-006-9019-0

Sbragaglia, V., Correia, R.A., Coco, S., Arlinghaus, R., 2019. Data
mining on YouTube reveals fisher group-specific harvesting patterns and
social engagement in recreational anglers and spearfishers. ICES J. Mar.
Sci. https://doi.org/10.1093/icesjms/fsz100

Schirpke, U., Meisch, C., Marsoner, T., Tappeiner, U., 2018. Revealing
spatial and temporal patterns of outdoor recreation in the European Alps
and their surroundings. Ecosyst. Serv. 31, 336--350.
https://doi.org/10.1016/J.ECOSER.2017.11.017

Selig, E.R., Hole, D.G., Allison, E.H., Arkema, K.K., McKinnon, M.C.,
Chu, J., de Sherbinin, A., Fisher, B., Glew, L., Holland, M.B., Ingram,
J.C., Rao, N.S., Russell, R.B., Srebotnjak, T., Teh, L.C.L., Troëng, S.,
Turner, W.R., Zvoleff, A., 2019. Mapping global human dependence on
marine ecosystems. Conserv. Lett. https://doi.org/10.1111/conl.12617

Stenseke, M., 2018. Connecting `relational values' and relational
landscape approaches. Curr. Opin. Environ. Sustain. 35, 82--88.
https://doi.org/10.1016/j.cosust.2018.10.025

Tenkanen, H., Di Minin, E., Heikinheimo, V., Hausmann, A., Herbst, M.,
Kajala, L., Toivonen, T., 2017. Instagram, Flickr, or Twitter: Assessing
the usability of social media data for visitor monitoring in protected
areas. Sci. Rep. 7, 17615. https://doi.org/10.1038/s41598-017-18007-4

Toivonen, T., Heikinheimo, V., Fink, C., Hausmann, A., Hiippala, T.,
Järv, O., Tenkanen, H., Di Minin, E., 2019. Social media data for
conservation science: A methodological overview. Biol. Conserv. 233,
298--315. https://doi.org/10.1016/j.biocon.2019.01.023

Varol, O., Ferrara, E., Davis, C.A., Menczer, F., Flammini, A., 2017.
Online Human-Bot Interactions: Detection, Estimation, and
Characterization. Elev. Int. AAAI Conf. Web Soc. Media.

Waldron, A., Mooers, A.O., Miller, D.C., Nibbelink, N., Redding, D.,
Kuhn, T.S., Roberts, J.T., Gittleman, J.L., 2013. Targeting global
conservation funding to limit immediate biodiversity declines. Proc.
Natl. Acad. Sci. 110, 12144--12148.
https://doi.org/10.1073/pnas.1221370110

Wood, S.A., Guerry, A.D., Silver, J.M., Lacayo, M., 2013. Using social
media to quantify nature-based tourism and recreation. Sci. Rep. 3.
https://doi.org/10.1038/srep02976
\end{scriptsize}

\appendix
\label{sec:Appendix}


\section*{Supplementary material}

# Assessment of CES from MPAs using social media analysis

1. Reviewer

2. Photo number

3. ID

    Unique ID number corresponding to the photo

4. Check box if picture is not valid

    *Check all that apply.*

    ☐ Not valid

5. MPA name

    *Mark only one oval.*

    ◯ galapagos

    ◯ rapanui

    ◯ great barrier reef

6. Brief description on photo content

_____

7. Pictures shows

*Mark only one oval.*

◯ A mostly natural landscape

◯ a mostly urban/ man-made landscape

8. Does the picture show:

*Check all that apply.*

☐ A purely terrestrial scene/topic?
☐ A purely marine/coastal scene/topic?
☐ Both scapes appear on the picture

Other: ☐ _____

9. If the picture is marine/coastal, does it represent an under or above water scene?

*Mark only one oval.*

◯ Underwater

◯ Above water

◯ Both

10. Cultural ecosystem service classification

*Check all that apply.*

- [ ] Artistic or cultural expressions and appreciation
- [ ] Living cultural heritage
- [ ] Gastronomy
- [ ] Historical monuments
- [ ] Landscape appreciation
- [ ] Nature appreciation
- [ ] Natural structures and monuments
- [ ] Religious, spiritual or ceremonial activities and monuments
- [ ] Research and education
- [ ] Social recreation
- [ ] Activity recreation
- [ ] Other

11. What is the main activity/benefit/value occurring in the photo (either appearing on picture or undertaken by photographer)

*Check all that apply.*

- [ ] Swimming
- [ ] Snorkelling
- [ ] Scuba diving
- [ ] Recreational fishing
- [ ] Artisanal fishing
- [ ] Wildlife watching
- [ ] Habitat appreciation
- [ ] Scientific
- [ ] Education
- [ ] Aesthetic (Landscape/Seascape)
- [ ] Kayak
- [ ] Other water sports
- [ ] Gastronomy
- [ ] Relaxation
- [ ] Spiritual
- [ ] Sunbathing, playing on the beach
- [ ] Heritage
- [ ] Art
- Other: [ ] _____

12. If there's wildlife in the picture: Classification of species

Please provide approximate description, e.g. dolphin, whale, marine turtle, marine lion, clown fish....write scientific name if known

_____

13. **Classification of species in picture**

    Please classify the species on the picture according to the categories below

    *Check all that apply.*

    - [ ] None present
    - [ ] Terrestrial mammals
    - [ ] Marine mammals
    - [ ] Reptiles
    - [ ] Bird species
    - [ ] Fish
    - [ ] Marine invertebrates
    - [ ] Terrestrial invertebrates
    - [ ] Elasmobranchs (sharks, rays...)

14. **Main habitat appearing on the picture**

    (habitat with greatest proportion of coverage in picture) (M = marine and coastal, T = terrestrial)

    *Check all that apply.*

    - [ ] NA
    - [ ] (M/T) Lagoon
    - [ ] (M) Shore
    - [ ] (M) Pelagic (water column)
    - [ ] (M) Coral
    - [ ] (M) Seagrass
    - [ ] (M) Rocky ground
    - [ ] (T) Rocky ground
    - [ ] (M) Sandy ground
    - [ ] (M) Macrophyte
    - [ ] (T) Humid
    - [ ] (T) Shrub
    - [ ] (T) Forest
    - [ ] Cliff
    - [ ] urban / Man-made

    Other: [ ] _____

# APPENDIX 2. GALAPAGOS AND EASTER ISLAND FIGURES

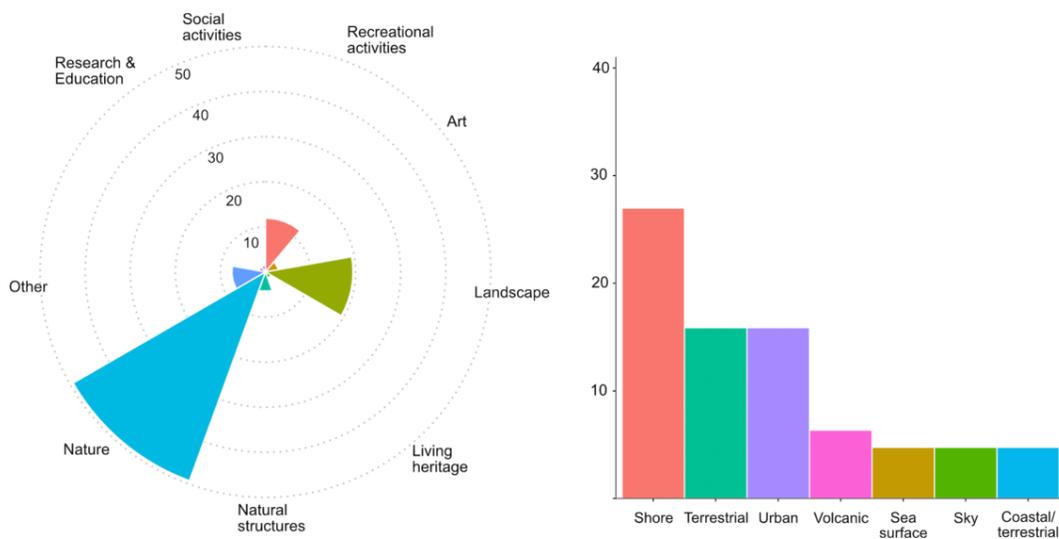

Figure 1. Instagram manual image content analysis for Galapagos. Left: percentage of photographs depicting specific CES. Right: percentage of photographs depicting specific habitats.

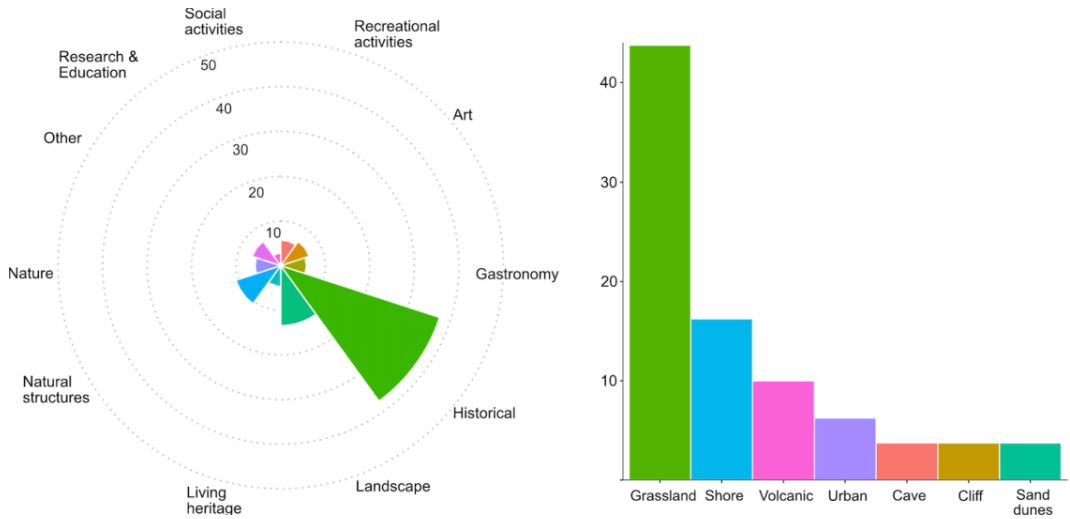

Figure. 2. Instagram manual image content analysis for Easter Island. Left: percentage of photographs depicting specific CES. Right: percentage of photographs depicting specific habitats.

Figure 3. Galapagos Instagram Eigenvector network

Figure 4. Hashtags communities generated through Fast Greedy algorithm from Galapagos Instagram Eigenvector network

Figure 5. Galapagos Twitter Eigenvector network

Figure 6. Hashtags communities generated through Fast Greedy algorithm from Galapagos Twitter Eigenvector network

Figure 7. Easter Island Instagram Eigenvector network

Figure 8. Hashtags communities generated through Fast Greedy algorithm from Easter Island Instagram Eigenvector network

Figure 9. Easter Island Twitter Eigenvector network

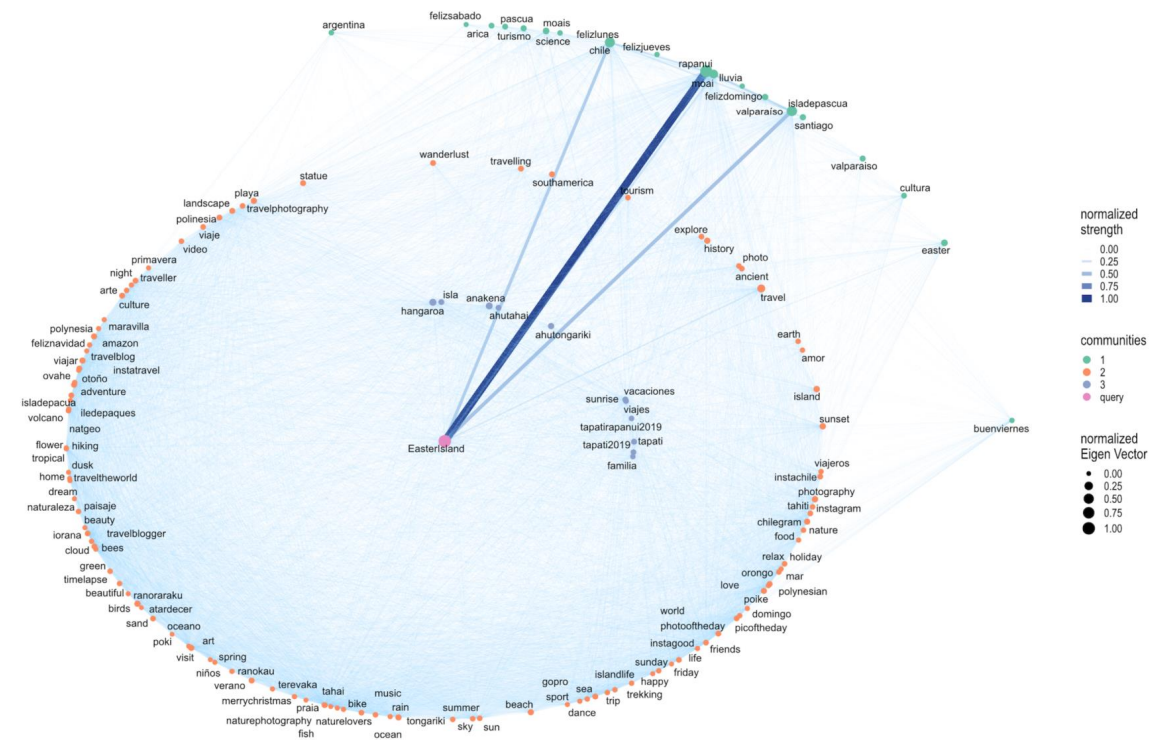

Figure 10. Hashtags communities generated through Fast Greedy algorithm from Easter Island Twitter Eigenvector network



\end{document}